\documentstyle[graphics]{aa}
\newcommand{\ls}
 {\mathrel{\hbox{\rlap{\hbox{\lower4pt\hbox{$\sim$}}}\hbox{$<$}}}}
\newcommand{\gs}
 {\mathrel{\hbox{\rlap{\hbox{\lower4pt\hbox{$\sim$}}}\hbox{$>$}}}}
\newcommand{\degg}{\hbox{$^\circ$}}
\newcommand{\arcm}{\hbox{$^\prime$}}
\newcommand{\arcs}{\hbox{$^{\prime\prime}$}}
\newcommand{\et}{et al.\ }

\newcommand{\xmm}{{\it XMM-Newton}}

\def\la{\mathrel{\hbox{\rlap{\hbox{\lower4pt\hbox{$\sim$}}}{\raise2pt\hbox{$
<$}}
}}}
\def\ga{\mathrel{\hbox{\rlap{\hbox{\lower4pt\hbox{$\sim$}}}{\raise2pt\hbox{$
>$}}
}}}

\begin{document}

\thesaurus{(11.01.2; 11.19.3: 13.25.2)}

\title{XMM-Newton First-Light Observations of the Hickson Galaxy Group 16}
\author{M.J.L. Turner\inst{1} 
\and J.N. Reeves\inst{1}
\and T.J. Ponman\inst{2}
\and M. Arnaud\inst{3}
\and M. Barbera\inst{16}
\and P.J. Bennie\inst{1}
\and M. Boer\inst{4}
\and U. Briel\inst{5}
\and I. Butler\inst{2}
\and J. Clavel\inst{6}
\and P. Dhez\inst{7}
\and F. Cordova\inst{8}
\and S. Dos Santos\inst{1}
\and P. Ferrando\inst{3}
\and S. Ghizzardi\inst{9}
\and C.V. Goodall\inst{2}
\and R.G. Griffiths\inst{1}
\and J.F. Hochedez\inst{15}
\and A.D. Holland \inst{1}
\and F. Jansen \inst{10}
\and E. Kendziorra\inst{11}
\and A. Lagostina \inst{9}
\and R. Laine\inst{12}
\and N. La Palombara\inst{9}
\and M. Lortholary\inst{3}
\and K.O. Mason\inst{13}
\and S. Molendi\inst{9}
\and C. Pigot\inst{3}
\and W. Priedhorsky\inst{14}
\and C. Reppin\inst{5}
\and R. Rothenflug\inst{3}
\and P. Salvetat\inst{15}
\and J. Sauvageot\inst{3}
\and D. Schmitt\inst{3}
\and S. Sembay\inst{1}
\and A. Short\inst{1}
\and L. Str\"{u}der\inst{5}
\and M. Trifoglio\inst{17}
\and J. Tr\"{u}mper\inst{5}
\and S. Vercellone\inst{9}
\and L. Vigroux\inst{3}
\and G. Villa\inst{9}
\and M. Ward\inst{1}}

\offprints{J.N Reeves}

\institute{X-ray Astronomy Group; Department of Physics and Astronomy;
Leicester University; Leicester LE1 7RH; U.K.
\and School of Physics and Astronomy, University of Birmingham, B15 2TT, UK.
\and CEA Saclay, 91191 Gif-sur-Yvette, France.
\and CESR Toulouse, BP 4346, 31028 Toulouse Cedex 4, France.
\and Max-Planck-Institut f{\"u}r extraterrestrische Physik, Postfach 1603,
85748 Garching, Germany
\and ESA-SOC, Vilspa, P.O.Box 50727, 28080, Madrid, Spain
\and LURE, Bat 209 D, Universite Paris Sud, 91405 Orsay, France.
\and Office of Research, University of California, Santa Barbara, CA93106, USA.
\and IFC Milan, 20133 Milano, Italy
\and PS Estec, Postbus 299, 2200 AG Noordwijk, Holland.
\and IAAP Tuebingen, D-72076, Germany
\and PX Estec, Postbus 299, 2200 AG Noordwijk, Holland.
\and MSSL Holmbury St Mary, Dorking RH5 6NT, UK
\and LANL, SST9, MS D436, Los Alamos, NM87545, USA
\and Institut d'Astrophysique Spatiale, Bat 121, Universite Paris Sud,
91405 Orsay, France
\and Osservatorio Astronomico di Palermo, Palermo 90134, Italy
\and ITESRE, 41010 Bologna, Italy}

\date{October 2000 / Accepted xxx 2000}

\maketitle

\begin{abstract}

This paper presents the \xmm\ first-light observations of the
Hickson-16 compact group of galaxies. Groups are 
possibly the oldest large-scale structures in the Universe, 
pre-dating clusters of galaxies, and are highly evolved. This group 
of small galaxies, at a redshift of 0.0132 (or 80 Mpc) is 
exceptional in the having the highest concentration of starburst or 
AGN activity in the nearby Universe. So it is a veritable laboratory 
for the study of the relationship between galaxy interactions and 
nuclear activity. Previous optical emission line studies 
indicated a strong ionising 
continuum in the galaxies, but its origin, whether from starbursts, 
or AGN, was unclear. Combined imaging and spectroscopy with the 
EPIC X-ray CCDs unequivocally reveals a heavily obscured AGN and a 
separately identified thermal (starburst) plasma, in NGC 835, NGC 
833, \& NGC 839. NGC 838 shows only starburst thermal emission. 
Starbursts and AGN can evidently coexist in members of this highly 
evolved system of merged and merging galaxies, implying a high 
probability for the formation of AGN as well as starbursts in 
post-merger galaxies.


\end{abstract}

\begin{figure*}
\resizebox{\hsize}{!}{\rotatebox{-90}{\includegraphics{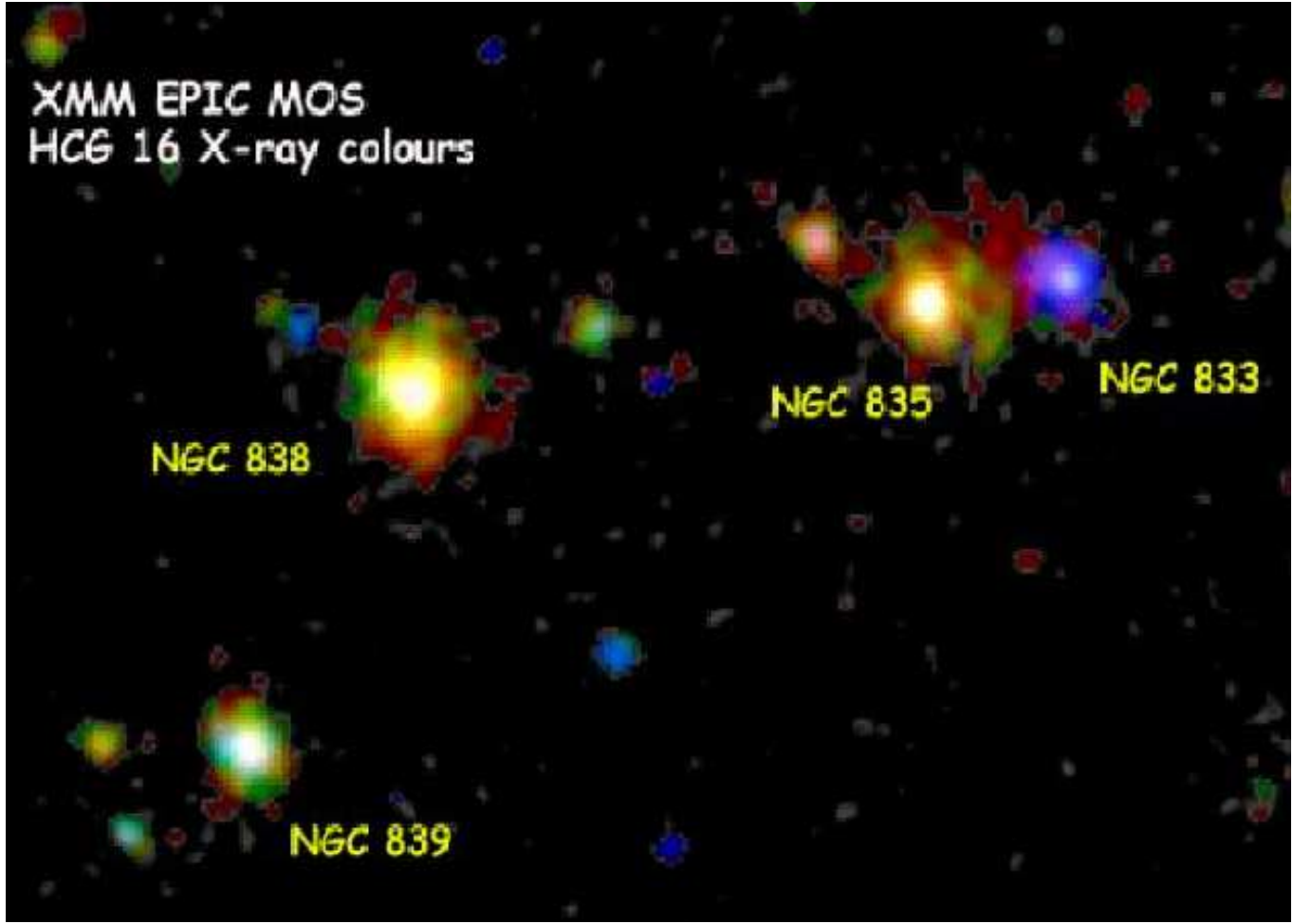}}}
\caption{The smoothed colour X-ray image of HGC 16 made with the EPIC 
MOS CCD imaging spectrometers on XMM-Newton. The spatial resolution 
is 6\arcs\ Full Width at Half Maximum, and 15\arcs\ Half Energy Width, limited 
by the mirrors. The energy band is 0.2-10 keV and the energy 
resolution varies from 140 eV FWHM at 6 keV to 70 eV at 500 eV. In 
the colour image, red corresponds to 800 eV and blue to $>3$~keV. The
physical scale across the image corresponds to 200~kpc (using 
$ H_0 = 50 $~km\,s$^{-1}$\,Mpc$^{-1}$). 
Notice the very hard (blue) nucleus of NGC 833 and the soft (red) halo
emission around the companion galaxy NGC 835.} 
\end{figure*}

\section{Introduction}

The Hickson-16 galaxy group (or HCG-16) comprises seven galaxies with a mean 
recession velocity (Ribeiro \et \cite{ribeiro}) of 3959$\pm$66km~s$^{-1}$ 
and a velocity dispersion of 86$\pm$55~km~s$^{-1}$, centered on the
position $\alpha=02\degg 09\arcm 33\arcs$, $\delta=-10\degg 09\arcm
46.7\arcs$ (J2000). The four 
central members of the group, originally identified by 
Hickson (Hickson \cite{hickson}), 
all fall within the 30\arcm\ field of view of EPIC,
they all show evidence for mergers and a strong ionising continuum; 
they have well resolved optical nuclei (Mendes de Oliveira \et
1998). However the nature of the ionising continuum is
unclear. Optical emission line diagnostics (de Carvalho \& Coziol
\cite{decarv},  Veilleux \& Osterbrock \cite{veil}) suggest 
that in NGC 838 and NGC 835 there is a starburst nucleus, whilst in NGC 
835, NGC833 and NGC 839 there is a low ionisation narrow emission 
line region (or LINER 2 nucleus). This could arise either from an AGN or from 
starburst activity: [OIII]/H$\beta$ values for the nuclei are all
$<2.5$ (de Carvalho \& Coziol \cite{decarv}). 
The EPIC instrument is able to observe directly the ionising 
continuum, and distinguish clearly between optically thin thermal emission
from a starburst, and non-thermal hard X-ray emission from an AGN. These 
observations can therefore be used to determine the nature of the 
ionising continuum, helping to clarify the relationship between 
mergers, the triggering of starbursts, and the creation and fueling 
of black holes.

\begin{figure*}
\resizebox{\hsize}{!}{\rotatebox{-90}{\includegraphics{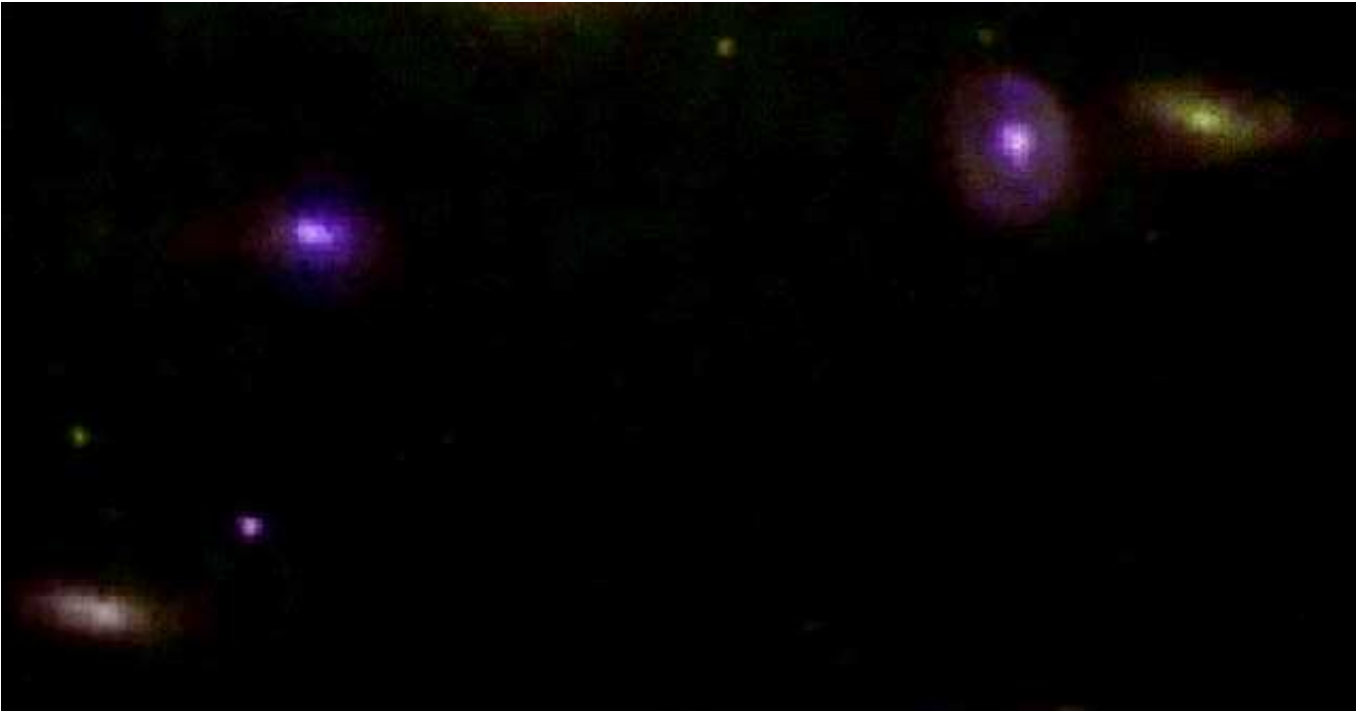}}}
\caption{The OM colour image of HCG 16 derived from 1000 second 
exposures in V and UV bands. A bright foreground star has been 
removed from the lower right of the image. Blue here represents the 
UV band.} 
\end{figure*}

\section{The XMM-Newton Observations}

The \xmm\ Observatory (Jansen \et \cite{jansen}) has three X-ray
telescopes of area $\sim1500$~cm$^{2}$, with the three EPIC instruments
at the foci; two of the EPIC 
imaging spectrometers use MOS CCDs (Turner \et \cite{turner}, Holland
\et \cite{holland})
and one uses a PN CCD (Str\"{u}der \et \cite{struder}). 
The observations of the HCG-16 galaxy group were taken in
orbit-23 as part of the \xmm\ EPIC first-light. 
Exposures of 50~ksec were taken with EPIC (sensitive from 0.2 
to 10 keV) and 1~ksec exposures 
were taken in V (550nm) and UV (280nm) with the XMM-Newton 
Optical/UV Monitor (OM) telescope (Mason \et \cite{mason}). 

The EPIC data were
processed using the pipeline scripts \textsc{emchain} (MOS) and
\textsc{epchain} (PN). Screening was applied using the \xmm\
SAS (Science Analysis Software). Hot and bad pixels and negative E3
events were removed from the data to reduce the level of electronic
noise. A low energy cut of 200~eV was
applied to the data. The first 10 ksec of data were also removed from the
EPIC observation, as this contained a high count-rate background
particle flare. The resultant exposure time for each of the detectors
was $\sim$40~ksec.

Figure 1 shows the resultant EPIC X-ray colour image of the centre of
the HCG-16 field. The hard, 
absorbed, spectrum of the AGN in NGC 833 shows up as a blue point 
source, and the soft starburst emission in the outer regions of NGC 
835 shows as a red halo; the other galaxies show extended X-ray 
discs. Figure 2 also shows the V-UV colour image from the OM. The nuclear 
regions of NGC 835 and NGC 838 show up brightly in the ultraviolet, 
indicative of hot stars or gas associated with enhanced star 
formation. There are also bright UV knots in the outer regions of NGC 
835 showing enhanced star formation there. A close-up OM image of NGC
835 is shown in figure 3. 

\begin{figure}
\resizebox{\hsize}{!}{\rotatebox{-90}{\includegraphics{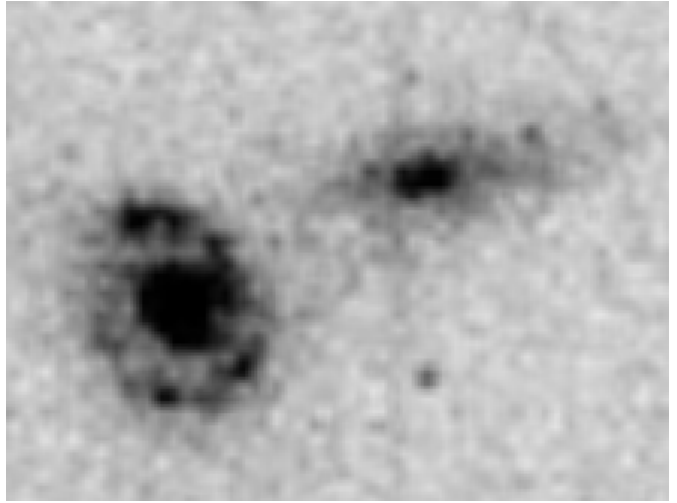}}}
\caption{A close-up, greyscale UV image from the Optical Monitor, showing the
galaxies NGC 833 (right) and NGC 835 (left). Bright UV knots, corresponding 
to possible regions of star formation, are seen in the outer disk of
NGC 835.} 
\end{figure}

\section{Spectral Analysis of the HCG-16 Galaxies}

Since EPIC resolves the optical disks of the 
galaxies, the spectra were prepared from photons falling within a 
region of interest based on the X-ray image. For NGC 833 the X-ray 
source is point-like, while for NGC 835 the core and the surrounding 
region (the red halo in figure 1) were analysed separately, 
the spectra of the remaining 
galaxies were made using the entire X-ray disc. 
Background spectra were taken from source-free regions
on the central EPIC-MOS and PN chips; the background spectra were 
normalised to the area of the source extraction regions. 

The background subtracted EPIC spectra 
were fitted, using \textsc{xspec v11.0}, with the latest response
matrices produced by the EPIC team; the systematic level of uncertainty 
is $<5$\%.  Finally spectra were binned to a minimum of 20 counts per
bin, in order to apply the $\chi^2$ minimisation technique. All 
subsequent errors are quoted to 90\% confidence 
($\Delta\chi^2=4.6$ for 2 interesting parameters). 
Values of $ H_0 = 50 $~km\,s$^{-1}$\,Mpc$^{-1}$ and $ q_0 = 0.5 $
have been assumed and all fit parameters are given in the rest-frame
of the HCG-16 system. We now present the individual EPIC spectra of the
4 main Hickson-16 galaxies. 

\subsection{NGC 833}

Optical imaging data on NGC 833 reveal a disturbed velocity field and 
pronounced misalignment of the kinematic and stellar axes, indicative 
of an ongoing interaction (Mendes de Oliveira \et \cite{mendes}). The emission
lines present in the optical spectra (de Carvalho \& Coziol \cite{decarv})
indicate weak non-stellar LINER-2 activity in the core; there is no optical 
evidence for current star formation ([NII]/H$\beta$$\sim$unity). 
The EPIC image of this galaxy is point-like, much smaller than the 
stellar disc.   

The best-fitting EPIC X-ray spectrum (Figure 4) shows three 
distinct components, all required at $>99.99$\% confidence. 
The most obvious is 
the peak at high energies from an obscured AGN; this emission is in 
the form of a power-law of index $\Gamma=1.8\pm0.5$, absorbed by material of 
column density equal to $N_{H}=2.4\pm0.4\times10^{23}$~cm$^{-2}$.  
The second component is an un-absorbed power-law, resulting from radiation 
scattered into our line of sight, by thin, hot, plasma directly 
illuminated by the AGN. The third component is radiation from an 
optically-thin thermal plasma, with a temperature of $kT=470$~eV. 
The improvement in the fit upon adding the thermal emission is
$\Delta\chi^{2}=36.7$.  A summary
of the fits to NGC 833 (and the other 3 galaxies) are given in table 1. 

This complex X-ray spectrum amply confirms the presence of an AGN in NGC 
833 of luminosity $1.4\pm0.6\times10^{42}$~erg~s$^{-1}$, it is, 
remarkably, the dominant source of power in the galaxy. In contrast, 
the thermal X-ray emission, is more than 100 times weaker 
($8.9\pm3.0\times10^{39}$~erg~s$^{-1}$) and the FIR luminosity
(Verdes-Montenegro \et \cite{verdes}) 
is also very low ($<3\times10^{42}$~erg~s$^{-1}$).

\begin{figure}
\resizebox{\hsize}{!}{\rotatebox{-90}{\includegraphics{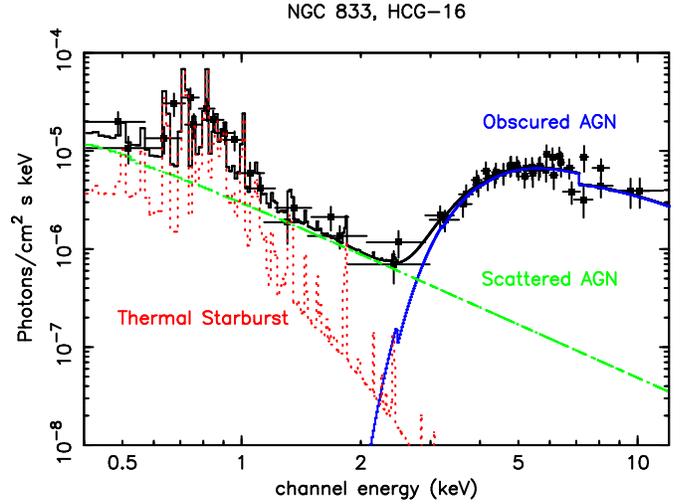}}}
\caption{The X-ray spectrum of the galaxy NGC 833. The most 
striking feature is the high-energy, absorbed power-law (at $>$~3keV) 
that is the direct emission from the active black hole at the centre 
of the galaxy. There is also an un-absorbed power-law, resulting from 
radiation scattered into our line of sight, by material 
directly illuminated by the AGN. These two spectral components 
together show the presence of an AGN of luminosity 
$1.4\pm0.6\times10^{42}$erg~s$^{-1}$. 
There is also weak soft X-ray emission from an optically thin plasma,
perhaps originating from starburst activity.}
\end{figure}

\subsection{NGC 835}

The adjacent galaxy, NGC 835 is undergoing a gravitational 
interaction with its neighbour NGC 833, as evidenced by the tidal 
tails in the optical image; and apparently contiguous stellar 
discs (Mendes de Oliveira \et \cite{mendes}). 
The velocity field is normal, but there is emission line 
evidence (de Carvalho \& Coziol \cite{decarv}) for LINER nuclear activity, 
and for current starburst 
activity in the outer regions; the knotted ring structure seen in the 
OM image supports this. 

\begin{figure}
\resizebox{\hsize}{!}{\rotatebox{-90}{\includegraphics{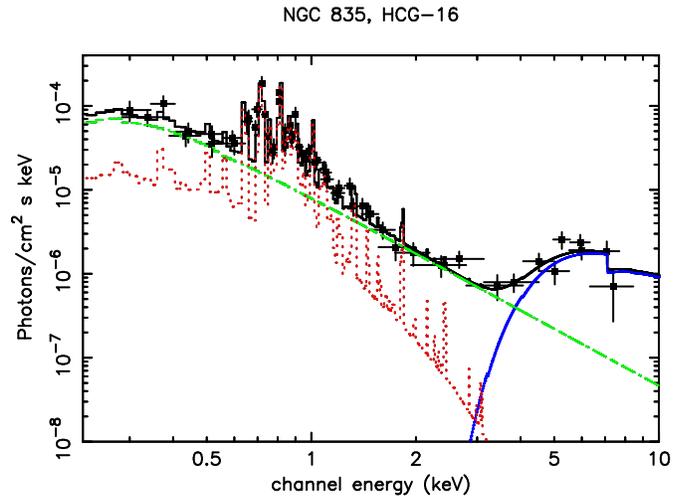}}}
\caption{The spectrum of the companion galaxy, NGC 835, 
which also shows an obscured active nucleus together with 
strong thermal soft X-ray emission. The AGN in both NGC 833 and NGC
835 may have been triggered by mutual gravitational interaction.}
\end{figure}

The X-ray emission from NGC 835 can be 
spatially separated into two areas, the core, and an outer region 
corresponding to the remainder of the stellar disc. The core has a 
very similar spectrum (Fig. 5) to that of NGC 833. There are 
absorbed and scattered power-laws indicating a heavily obscured AGN 
($N_{H}=4.6\pm1.5\times10^{23}$~cm$^{-2}$) of luminosity 
$1.2\times10^{42}$~erg~s$^{-1}$ 
(0.5-10 keV), but the soft thermal component is more luminous than NGC 
833 at $2.5\pm0.3\times10^{40}$~erg~s$^{-1}$; 
it is almost certainly from current starburst activity 
and the FIR luminosity (Verdes-Montenegro \et \cite{verdes}) 
is 100 times larger at 
$2.7\times10^{44}$~erg~s$^{-1}$. The spectrum at the periphery of NGC 835 is 
purely thermal, with a temperature of 300 eV and a luminosity of 
$2.9\pm0.7\times10^{40}$~erg~s$^{-1}$, similar to that of the core. 
This is the 
X-ray emission from the starburst region including the ring structure 
and knots seen in the OM V-UV image. 

To summarize, the EPIC data clearly show the presence of an AGN, in
both NGC 833 and NGC 835, that coexists with present starburst activity 
in the core (and for NGC 835 in the periphery) of the galaxies. The detections
of the obscured AGN and thermal starburst components in both galaxies 
are highly significant, at $>99.99$\% confidence (see table 1). 

\begin{table*}
\centering
\caption{X-ray spectral fits to the 4 HCG-16 galaxies. $^a$ Temperature
of thermal component in keV. $^b$ Column density of the absorbed power-law
in units of 10$^{22}$~cm$^{-2}$. $^c$ Improvement in the spectral fit
upon adding the obscured hard power-law. $^d$ Improvement in the fit
from adding a soft, thermal (Mekal) component. $^e$ Best-fit reduced
chi-squared. $^f$ Indicates that parameter is fixed. Note ABS.PL is
the absorbed power-law, SCAT.PL is the scattered power-law; $\Gamma$
for these 2 components have been tied.}

\begin{tabular}{@{}lllccccc@{}}
\hline                 

\ & \ & \multicolumn{3}{c}{Hard Component} & \multicolumn{2}{c}{Thermal
Soft Component} \\

Galaxy & Model & $\Gamma$ or kT$^a$ & $N_H$$^b$ & $\Delta\chi^2$$^c$ &
kT$^a$ & $\Delta\chi^2$$^d$ & $\chi_{\nu}^{2}$$^e$ \\

\hline

NGC 833 & ABS.PL + SCAT.PL + MEKAL & $\Gamma=1.8\pm0.5$ & 24$\pm$4 & 92.7 &
0.47$\pm$0.12 & 36.7 & 0.632 \\

NGC 835 (centre) &  ABS.PL + SCAT.PL + MEKAL & $\Gamma=2.25\pm0.23$ &
46$\pm$15 & 58.4 & 0.51$\pm$0.07 & 176.2 & 1.02 \\

NGC 835 (diffuse) & MEKAL $\times2$ & kT=4~keV$^f$ & -- & -- &
0.31$\pm$0.05 & 73.4 & 1.2 \\

NGC 838 & MEKAL $\times2$ & kT=3.2$\pm$0.8 & -- & -- & 0.59$\pm0.04$ &
209.5 & 1.21 \\

NGC 839 & ABS.PL + SCAT.PL + MEKAL & $\Gamma=2.1\pm0.8$ & 45$\pm$20 &
12.0 & 0.63$\pm$0.10 & 40.9 & 1.38 \\

\hline
\end{tabular}
\end{table*}

\subsection{NGC 838}

NGC 838 is an ongoing merger with strong starburst activity. 
Optical data (Mendes de Oliveira \et \cite{mendes}) show kinematic warping, 
and multiple velocity components in the ionised gas, and a double 
optical core (also see de Carvalho \& Coziol \cite{decarv}). The 
infrared luminosity is $3.3\pm10^{44}$~erg~s$^{-1}$. The EPIC spectrum of 
NGC 838 (Fig. 6) shows purely thermal emission, the disc is resolved in 
X-rays, but there is no separate sharp core in the X-ray image. The 
spectrum is fitted with a two temperature thermal spectrum ($kT=3.2$~keV 
and $kT=590$~eV) and the luminosity is high at 
$1.9\pm0.3\times10^{41}$~erg~s$^{-1}$. 
This is all consistent with the optical data: the soft X-ray 
emission is from the ionised gas produced in the starburst while the 
hard thermal spectrum could be characteristic of unresolved X-ray 
binaries. There is no statistically significant scattered or obscured 
power law; the upper limit for the AGN luminosity is 
$5\times10^{40}$~erg~s$^{-1}$, assuming an absorbing column of 
$5\times10^{23}$~cm$^{-2}$.

\begin{figure}
\resizebox{\hsize}{!}{\rotatebox{-90}{\includegraphics{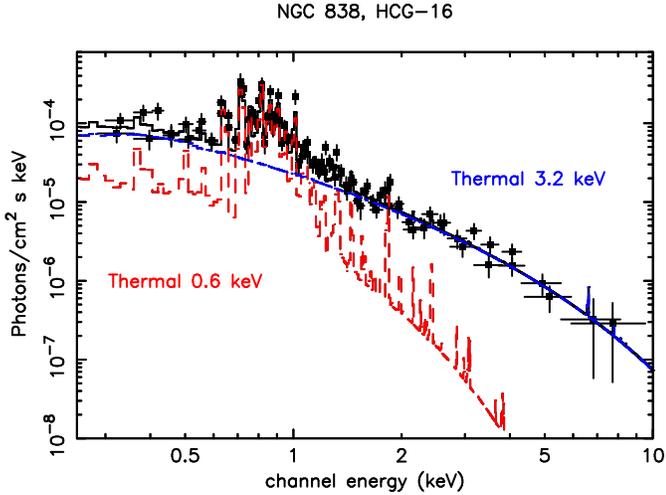}}}
\caption{The EPIC-MOS spectrum of NGC 838. Only emission from the 
starburst is present, with no detectable hard X-ray emission from a 
central AGN; the hard X-ray emission could arise from unresolved 
X-ray binaries in the galaxy.}
\end{figure}

\subsection{NGC 839}

\begin{figure}
\resizebox{\hsize}{!}{\rotatebox{-90}{\includegraphics{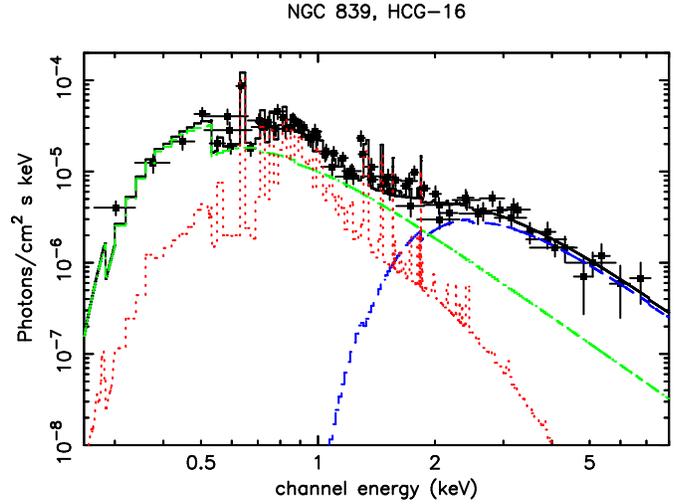}}}
\caption{The X-ray spectrum of NGC 839. There is both 
starburst emission and emission from a low luminosity obscured AGN. 
Unusually the elemental abundance of NGC 839 appears to be $\sim$5 
times solar.}
\end{figure}

NGC 839 may also be a recent merger (Mendes de Oliveira \et \cite{mendes}, 
de Carvalho \& Coziol \cite{decarv}), it has a double nucleus in the 
optical, a FIR luminosity of $3.1\times10^{44}$~erg~s$^{-1}$, and a disturbed 
velocity field; optical lines indicate an active LINER-2 nucleus (de
Carvalho \& Coziol \cite{decarv}). 
In the soft X-ray EPIC spectrum (Fig. 7) there is
optically-thin thermal emission, 
similar to that of the other galaxies, of temperature $kT=600$~eV, and 
luminosity $1.8\pm0.3\times10^{40}$~erg~s$^{-1}$; 
a typical indicator of a current starburst. 
The spectrum also shows an obscured AGN, as found in NGC 
835 \& NGC 833; it is however much less luminous
($8\pm3\times10^{40}$~erg~s$^{-1}$) for a column of
$N_{H}=5\times10^{23}$~cm$^{-2}$. Interestingly the abundances in NGC
839 appear to be higher than solar. There are apparent weak
Lyman-$\alpha$ lines of O, Mg and Si in the EPIC spectrum, although 
the significance of
these features is low (at only 90\% confidence). Fitting the soft X-ray
spectrum with the \textsc{mekal} model does however yield an over-abundance of
5.2$\pm$2.0 times the solar value. One interesting possibility is
that the heavier elements have been enriched through the intense
starburst activity in this galaxy.  

\section{Conclusions}

Direct X-ray spectroscopy is the best way to identify hidden AGN in 
galaxies, and here the EPIC cameras on XMM-Newton have produced clear 
evidence for active, massive black holes in three out of four 
galaxies in HCG-16. The presence of a similar active nucleus in NGC 838 
is unlikely, unless it is very heavily absorbed. The nature of the 
ionising continuum in the four galaxies has been elucidated: there is 
thermal emission from starburst activity in three, (possibly four) of 
the galaxies, and in three of them there is a coexisting active black 
hole. While LINER-1 galaxies with broad H$\beta$ lines do harbour black 
holes (Terashima \et \cite{tera-one}, \cite{tera-two}) 
this is the first {\it direct} evidence that black holes power 
LINER2 galaxies. These AGN are at the low end of the luminosity 
scale, consistent with their small size (Magorrian \et \cite{mag}). 

In NGC 833, the accreting black hole X-ray luminosity arguably 
exceeds the FIR luminosity. This 
is very unusual, even compared with much more luminous AGN, and may 
indicate the stripping of dust and gas by past interactions with 
other galaxies (Mendes de Oliveira \et \cite{mendes}). The EPIC X-ray study of 
this nearby and evolved system of small galaxies indicates a high
fraction of active black holes coexisting with starbursts. This
is consistent with optical studies of compact groups 
(e.g. Coziol \et \cite{coziol}), 
where a large fraction of galaxies with nuclear activity is found. 
The observations may also provide a link between normal 
galaxies, where black holes may be inactive, and Seyfert galaxies and 
quasars, where black holes dominate, and where galaxy mergers may be 
implicated in the onset of black hole activity (Boyce \et \cite{boyce}, 
Bahcall \et \cite{bahcall}). Further 
observations with XMM-Newton have the potential to determine more 
precisely the fraction of nearby galaxies that harbour active, 
low-luminosity black holes. 

\section*{ Acknowledgements }

This work is based on observations obtained with XMM-Newton, an ESA science 
mission with instruments and contributions directly funded by 
ESA Member States and the USA (NASA). 
EPIC was developed by the EPIC Consortium led by the Principal 
Investigator, Dr. M. J. L. Turner. The consortium comprises the 
following Institutes: University of Leicester, University of 
Birmingham, (UK); CEA/Saclay, IAS Orsay, CESR Toulouse, (France); 
IAAP Tuebingen, MPE Garching, (Germany); IFC Milan, ITESRE Bologna, 
OAPA Palermo, (Italy). EPIC is funded by: PPARC, CEA, CNES, DLR and
ASI. Finally we thank the referee, Reinaldo de Carvalho, for his
report, and for some useful comments and suggestions.

%\vspace{-0.5cm}

\label{lastpage}

\end{document}